\documentclass{iopart}
\usepackage {amsfonts}
\usepackage {graphicx}
\usepackage{iopams}
\newcommand{\Vmeas}{ {\mathbf{V}_{\mbox{\tiny meas}}}}
\newcommand{\srefer}{ {\mathbf{s}_{\mbox{\tiny ref}}}}
\begin{document}

\title{EIT Reconstruction Algorithms: Pitfalls, Challenges and Recent Developments}

\author{ William
R.B. Lionheart}

\address{ Department of Mathematics, UMIST, UK}

\begin{abstract}
We review developments, issues and challenges in Electrical
Impedance Tomography (EIT), for the 4th Conference on Biomedical
Applications of EIT, Manchester 2003. We focus on the necessity for
three dimensional data collection and reconstruction, efficient
solution of the forward problem and present and future reconstruction
algorithms. We also suggest common pitfalls or ``inverse crimes'' to avoid.

{\em This paper is dedicated to the memory of the author's
 Mother, Sheila Breckon, 1926-2003}
\end{abstract}

\section{Introduction}
As the papers in this special issue and the 4th Conference on
Biomedical Applications of EIT show, reconstruction algorithms for
electrical impedance tomography remain an active and exciting area  of
research. In this article we aim to draw attention to the best current
practice as well to pitfalls to avoid, also to highlight some active
areas of development and some of the promising methods that have not yet
been implemented in practical algorithms. We hope this paper will be a
useful contribution for those new to the field as well as being thought
provoking for those of us who have been working in the area for some
time. Like any review this article has a personal slant on the subject,
it also fails to be either a comprehensive review or a history of the
subject. In some areas we have deliberately provided less detail where 
there accessible articles covering the material, in this volume or
elsewhere.


By Electrical Impedance Tomography (EIT) we mean the process of
estimating internal admittivity (complex conductivity) changes from
low frequency current and voltage measurements through a system of
electrodes at the surface. Although the details of the instrumentation
design varies between biomedical, geophysical and industrial
applications the mathematics of the reconstruction problem is
essentially the same. The main differences being in the configuration
of the electrodes used, and the {\em a  priori} information available.

\section{Three dimensionality}

In most cases where electrical imaging is used the problem is
genuinely three-dimensional. An  exception being industrial or
non-destructive testing problems where the conductivity and the
electrodes can be assumed
independent of one  coordinate. Why then is there such a large body of
published work in which data is collected on a three dimensional body
and the reconstruction performed assuming the body was two
dimensional? There are two main reasons: speed and the fact that to
some limited extent it worked. The issues of speed include the speed
of data collection. The belief that electrodes arranged in a
single plane would yield an adequate reconstruction of the
conductivity in that plane, and consequent use of a small number of
electrodes resulted in a fast data collection system. The number of
electrodes applied is also a significant factor when they have to be
applied individually to the skin which in itself is a time consuming
process. However it is known that data measured on a three dimensional
body cannot be fitted accurately to any two dimensional conductivity
distribution \cite{LionhShDim}. Moreover attempts to fit a two dimensional model result
in errors of position and shape of anomalies \cite{GobleTh, PaiviTh}. Another factor in
the choice of two dimensional reconstruction methods was the cost of
fast processors and memory required to perform three-dimensional
forward modelling and reconstruction, which in the 1980s and early
1990s were prohibitive.

Of these factors the one which remains is the inconvenience of
applying a fully three dimensional system (for example multiple
planes) of electrodes to a human subject, a technological problem
which must be overcome if EIT is to be used effectively as a medical
imaging technique.

\subsection{Is it tomography?}
It is worth mentioning here that EIT, despite its now traditional name
agreed at the first Sheffield EIT meeting in 1986, is not {\em
tomographic} in that it is not possible to reconstruct an image
slice-by-slice. (One can of course reconstruct a three dimensional conductivity
distribution and then display it on any desired slice.) The physical
explanation for this is that, unlike X-rays, low frequency electric
current cannot be confined to a plane even by a system of external
electrodes, and that a change in conductivity anywhere in the domain
can affect all measurements not just those on a ray path.

An obvious extension of the traditional array of equally spaced
electrodes in a plane to a three dimensional data collection system is
to employ multiple planes of electrodes. If electrodes are excited in
pairs on a given plane it is necessary to make measurements of voltage
on the electrodes
in other planes as well. If one is modifying the traditional planar
system for this purpose, and voltages are measured between adjacent
pairs of electrodes, one must also make measurements between planes as
well for a complete set of transfer impedance data. Clearly only one
measurement between one pair of electrodes on adjacent planes is
sufficient by superposition. This configuration is however only an
expedient way to employ inflexible equipment and is not likely to be
an ideal data collection scheme. Optimal current drive
patterns have been described for a cylindrical tank  with
several planes of electrodes~\cite{GobleOpt}.
The optimal arrangement of electrodes
for imaging brain or lung function for example has yet to be determined.

There is interesting  connection between EIT and tomographic imaging
which has yet to be exploited in a practical reconstruction algorithm.
Suppose that one is able to measure the complete transfer impedance on
a plane intersecting a three dimensional object, while there is no
known way of reconstructing the conductivity on this plane Greenleaf
and Uhlmann \cite{GU} show that the integral of the conductivity over
that plane is known. Suppose now that we use a large number of surface
electrodes so that the transfer impedance, and hence the integral of
the conductivity, is known for a wide class of planes. Reconstructing
the conductivity reduces to the problem of inverting a Radon plane transform
with limited data.

\section{Forward Problem solution}

For accurate EIT reconstruction it a prerequisite to have a model
capable of predicting the voltages on electrodes for a given
conductivity distribution. The conductivity is adjusted until the
voltages fit to measurement precision. This forward model must also be
capable of predicting the electric fields in the interior given the
conductivity, as this is used in the reconstruction algorithms (See
 \sref{sec:jac}). It is
clear that the voltages on the electrodes must be predicted to a
precision better than the accuracy of the measurements. But the
accuracy required of the  interior electric fields and consequently
the sensitivity of measurements to conductivity changes, is a matter
which has received little attention.

\subsection{Choice of method}
As we are interested in non-homogeneous conductivity distributions on
irregular domains Finite Element Method (FEM) is the natural
choice~\cite{YorkeyTh,BreckonTh,MarkoTh,Marko,PaiviTh, NickTh,Nick,Marc} ,
although finite difference~\cite{Patterson} and finite volume methods ~\cite{Dong,Zlochiver}have also been
employed. Where the conductivity is known and homogeneous in some sub
domain, especially a neighbourhood of the boundary, and attractive
proposition is to use a hybrid boundary element and finite element
method~\cite{Hsiao}. As the highest field strengths are typically near  the edges
of electrodes this is where the densest mesh is required. With the
Boundary Element Method (BEM) only the surface needs to be discretized with a
consequent saving on the size of matrix to be inverted. However BEM
alone can only be used on homogeneous regions, and is useful for
fitting impedance values to known regions~\cite{deMonk} . On the other
hand the matrix for the hybrid system has a dense blocks  for
the BEM and for the coupling between BEM and FEM, and care has to be taken
to use an efficient solution scheme for such a matrix. 

\subsection{Custom versus commercial FEM code}
For medical applications, as well as industrial applications where
conductive electrodes are in contact with an aqueous solution, the
Complete Electrode Model must be employed for accurate prediction of
the electrode voltages \cite{CEM}. This takes account of both the shunting
 effect
of the conducting electrodes and the contact impedance layer between
the electrode and the solution (see \eref{eq:CEM} below).  As this is not a
 standard type of
boundary condition it is not easily implemented in commercial finite
element packages where the source code is not accessible. This
together with the necessity to update the system matrix rapidly with a
revised conductivity militates against the use of a closed source 
commercial FEM solver. Another factor is the ease with which the
Jacobian can be assembled from low level access to products of
gradients of shape functions (see \sref{sec:jac}). Even FEMLAB \cite{COMSOL}
 which has considerable
low-level access to internal structures through Matlab does not allow
this at present. 

In practice the assembly of the FEM system matrix, at least for simple
linear tetrahedral  elements, is easily implemented  and the step from two to
 three
dimensions requires only the calculations of integrals of products of
shape functions over faces under electrodes \cite{PaiviTh,NickTh}. The steps
 which involve
considerably more time and care are mesh generation and the efficient
solution of the linear system. The design of three dimensional mesh
generators is a major research topic in itself, and we have yet to
find an existing program, commercial or free, which is ideal for EIT.
The main requirements are for an efficient mesh of an object composed
of smooth but irregular surfaces, which respects interior boundaries,
and electrodes on the surface. The mesh density needs to be determined
as a function of position so that high field strengths (for example
near electrodes) can be accurately represented without excessive
density in areas where the field varies slowly. Even if such a program
were available one would still need to measure the external shape of
the body accurately, and in the case of the human head intricate
internal structures such as the skull needs to be segmented from X-Ray
CT or MRI scans \cite{Bayford}.

\subsection{Efficient forward solution}
The computational complexity of the three-dimensional forward problem
means that attention has to be paid to the efficiency of the forward
solution. For two dimensional problems a standard approach to solving
the linear system resulting from the finite element discretization is
to employ Cholesky factorization (or $LU$ decomposition for the case of
complex admittivity) together with forward and backward substitution
to give the potential for each applied current pattern \cite{GolubvanL}. As the
 system
matrix is sparse, renumbering the degrees of freedom to  reduce the
fill in of non-zeros in the Cholesky factors is common practice. The Symmetric
Multiple Minimum Degree algorithm is a standard choice which tends to
move degrees of freedom with higher ``valencies'' (coupled to many
others) to the bottom of the vector. With the complete electrode model
degrees of freedom include both finite element nodal values and
electrode voltages. The Matlab function {\tt symmmd} implements the
multiple minimum degree algorithm, and for the complex case the Column
Multiple minimum degree {\tt colmmd} function can be used which
reduces the fill in for the $LU$ factorization. It needs to be
stressed that although node renumbering can be expensive it needs to
be done only once for each mesh and the result stored for future use.
In three dimensions
iterative solution methods have become more attractive, although each
iterative step must be applied to the multiple right-hand sides for
each current pattern. Iterative methods also have an advantage that
for when the update to the conductivity is small the change in the
potential will also be small. This means that the number of iterations
need to solve the forward solution can be reduced as the iteration can
be initialized with the previously computed values. 

For the real case Conjugate Gradient method is a
favourite choice. Convergence is improved using a preconditioner which
is an approximate inverse to the system matrix. If the conductivity
does not vary over too wide a range during the reconstruction process
this approximate inverse can be precomputed based on a typical
conductivity, and as the inversion is typically performed by
incomplete Cholesky factorization the same consideration of node
renumbering applies as for direct methods. For the complex case other
Krylov subspace iterative methods \cite{NickKS} such as GMRES can be
used \cite{GMRES}.

\subsection{Accuracy of forward solution}
In iterative methods time can be saved if a solution is not required
to full machine precision, and as our boundary voltage data is not so
accurate it is unnecessary to solve to this accuracy. The accuracy
required for the interior electric fields and hence the Jacobian is
not well understood. That said the accuracy of the finite element
method approximation is well studied and there are {\em a priori} estimates
\cite{Babuska} for the error in terms of the mesh size ($r$ convergence)
and the order of the elements used ($h$ convergence). There are also
{\em a posteriori} error estimates based on calculated solutions \cite{Babuska}
 which have not yet been widely used in EIT. Although there
is some work using higher order elements \cite{PaiviTh} the best choice
of element for EIT remains an open problem, and the possibility of
using vector elements \cite{Bossavit} to calculate electric fields and current
densities accurately in the interior is largely unexplored for our
problem. The use of infinite elements to model unbounded regions, or
at least regions which while bounded have a substantial part where we
have no surface data, is an interesting possibility~\cite{PaiviInf}. Possible
 applications include limbs when the torso is being imaged, as
well as problems where the body is treated as an infinite half space
such as in geophysical imaging or the use of a small surface electrode
array in medical EIT.

\subsection{Multigrid and adaptive meshing}
So far we have considered the situation in which the same mesh is used
throughout the forward solution process. One attractive alternative is
to solve first on a course grid to give the crude features of a
solution which is then extrapolated to a finer grid where a mode
accurate solution is calculated. A systematic treatment of this
cycling between finer and courser grids is used in multigrid  solution
algorithms. At least for regular grids the theoretical optimum of
solving a system with time complexity $O(N)$ can be approached for
$N$ degrees of freedom. The geometric problems of calculating a
hierarchy of grids for a complex object can be avoided using algebraic
multigrid methods which are described elsewhere in this special issue \cite{ManuchCath}.

Another strategy for reducing the solution time is to adaptively vary
the finite element mesh used. During the forward solution process for
a fixed conductivity the mesh density is increased where the high
field strengths are found and decreased where the potential is gently
varying. This results in a more accurate solution than a regular mesh
with the same number of degrees of freedom. One complexity for our
problem is that we have multiple right-hand sides, so we can choose
either to use the same mesh for all current patterns, in which case it may
be unnecessarily fine near passive electrodes, or to use a different mesh for
each drive configuration. Although typically the highest field
strengths appear near the edges of electrodes (including passive
electrodes if the contact impedance is low), sharp contrasts in
conductivity can also give rise to high field strengths. For example
an aperture in the skull (low conductivity) for a relatively high
conductivity blood vessel or nerve. An adaptive meshing algorithm will
increase the mesh density here once the conductivity contrast has been
predicted by the reconstruction algorithm. The thesis of Mollinari~\cite{Marc}
explores the use of adaptive meshing in both forward and inverse
problems in EIT and gives a clear indication that this is a fruitful
area for further study.

\section{Regularised Newton Methods}

While numerous {\em ad hoc} reconstruction methods have been tried for
EIT the standard approach is to use one of a family of regularized
Newton-type methods. The approach is to some extent the obvious one:
the problem is non-linear so linearize, the linear problem is
ill-posed so regularize, the linear approximation cannot
reconstruct large contrasts or complex geometries so the process must
be applied iteratively. There are of course many variations on this
basic approach and we will sketch some typical ones. First let us
assume that the conductivity $\sigma$ has been represented by a finite
number of parameters $\mathbf{s}$. In the simplest case,  this is taken  as a
sum of basis functions such as the characteristic functions of a set
of regular or irregular voxels, or smooth basis functions. Other
choices would include a detailed model involving conductivity values as
well as parameters describing shape internal regions \cite{Ville1,Ville2}.

\subsection{Regularization}
 Our
forward operator $F$ gives us $\mathbf{V}=F(\mathbf{s})$ the simulated voltages
at the boundary. We will leave aside the adaptive current approach
\cite{GIN,BandP} where the current patterns used depend on the estimated
 conductivity. As the
goal is to fit the voltage measurements $\Vmeas$, the simplest approach is to
minimize the sum of squares error 
\begin{equation} || \Vmeas - F(\mathbf{s})||^2\end{equation}
the so called {\em output least squares} approach. Here $\|\cdot\|$ is
the standard 2-norm on vectors. In
practice it is not usual to use the raw least squares approach, but at
least a weighted sum of squares which reflects the reliability of each
voltage. Such approaches are common both in  optimization  and 
the statistical approach to inverse problems.

Minimization of the voltage error (for simple parameterizations
of $\sigma$) is doomed to failure as the
problem is illposed~\cite{VauhkonenBasis}. In practice the minimum lies in a long narrow
valley of the objective function\cite{BreckonTh}. For a unique
solution one must include additional information about the
conductivity, an example is to include a penalty $G(\mathbf{s})$ for highly
 oscillatory
conductivites, hence  in our problem we seek to minimize
\begin{equation}\label{eq:f}f(\mathbf{s})= || \Vmeas - F(\mathbf{s})||^2+ G(\mathbf{s}).\end{equation}
A typical simple choice \cite{MarkoTh} is 
\begin{equation}\label{eq:GL}G(\mathbf{s}) = \alpha^2||L(\mathbf{s}-\srefer)||^2\end{equation}
 where
$L$ is a matrix approximation to some partial differential operator
and $\srefer$ a reference conductivity (for example
including known anatomical features).  The minimisation of $f$
represents a trade-off between fitting the data exactly and not making
the derivatives of $\sigma$ too large, the trade off being controlled
by the regularization parameter $\alpha$. Other smooth choices of $G$ include
 the inverse of a Gaussian
smoothing filter~\cite{AndreaTMI}. In these cases where $G$ is smooth
and for $\alpha$ large enough the Hessian of $f$ will be positive
definite, we can then deduce that $f$ is a {\em convex}
function~\cite[Ch 2]{Vogel}, so that a critical point will be a strict
local minimum, guaranteeing the success of smooth optimization
methods. Such regularization however will prevent us from
reconstructing conductivities with a sharp transition, such as an
organ boundary. Including the {\em Total Variation}, that is the
integral of $|\nabla{\sigma}|$, still rules out wild fluctuations in
conductivity while allowing step changes. The cost is that the
inclusion of an absolute value destroys the differentiability of $f$
and means that we must employ non-smooth optimization
methods. This is both more difficult and computationally expensive
than smooth optimization problems. Early applications to
EIT~\cite{DobsonSantosa} employed total variation regularization to
the linearized EIT problem. A detailed review of the literature on
this subject as well as a more efficient optimization algorithm can be
found in the thesis~\cite{BorsicTh}.
See~\cite{WadeSenior} for analysis of Total Variation
regularization of EIT.

In the
statistical approach \cite[Ch 4]{Vogel} to regularization the minimizer of $f$
maximizes the {\em a posteriori} probability (the {\em MAP estimate}) assuming
 independent
Gaussian error with mean zero and unit variance  on the measurements and
the {\em a priori} information on $\sigma$ represented by the
probability distribution $\exp( - G(\mathbf{s})/2)$. To many the
probabilistic  approach to regularization provides a more rational
frame work for the recovery of a finite number of parameters from a
discrete set of data than an approach  to regularization based on
functional analysis. Ideally we would incorporate a probability model
for the errors in the data and using a prior probability density for
the unknown parameters find not just the maximum of the posterior
probability density, but a more compete description of the probability of
a range of conductivity images. For an excellent review of the
statistical regularization in EIT see~\cite{Kaipio}. Despite the low
spatial resolution of EIT the temporal resolution is high. In
biomedical applications to time varying conductivity distributions, 
caused for example by blood or air flow, images are correlated
temporally as well as spatially. Statistical time series methods such
can be used to include both spatial and temporal
regularization~\cite{VauhkonenKarl,VauhkonenFixed}. Applications to
industrial process modelling ~\cite{Seppanen} are better developed than biomedical
applications, although this is a highly active area of research.

\subsection{Linearized Problem}
Consider the  simplified case  where $F(\mathbf{s}) $ is replaced by a linear
 approximation 
\begin{equation} \label{eq:linF} F(\mathbf{s}_0) + J(\mathbf{s}-\mathbf{s}_0) \end{equation}
where $J$ is the Jacobian matrix of $F$ calculated at some initial
conductivity estimate $\mathbf{s}_0$ (not necessarily the same as $\srefer$).
The function to be minimized~\eref{eq:f} with regularizing penalty
term~\eref{eq:GL} becomes a quadratic function when $F$ is replaced by its
linear approximation~\eref{eq:linF}.
 Defining $\delta\mathbf{s}=
\mathbf{s}-\mathbf{s}_0$ and $\delta\mathbf{V} = \Vmeas -
F(\mathbf{s}_0)$
the solution to the linearized regularization problem is given by 
\begin{equation} \label{eq:Tik}\delta\mathbf{s} =(J^*J+ \alpha^2
 L^*L)^{-1}(J^*\delta\mathbf{V}+\alpha^2L^*L(\srefer-\mathbf{s}_0))
 \end{equation}
(where $J^*$ is the conjugate transpose of $J$) or any of the equivalent forms \cite{Tarantola}. While there are many
other forms of regularization possible  for a linear ill-conditioned
problem this generalized Tikhonov (or Tikhonov-Phillips)
regularization has the benefit that the {\em a priori} information it
incorporates is made explicit and that under Gaussian assumptions it
is the statistically defensible MAP estimate. If only a linearised
solution is to be used with a fixed initial estimate $\mathbf{s}_0$ the
Jacobian $J$ and a factorization of $(J^*J+ +\alpha^2 L^*L)$ can be
precalculated off-line. The efficiency of this calculation is then
immaterial and the regularized solution can be calculated using the
factorization with complexity $O(N^2)$ for $N$ degrees of freedom in
the conductivity (which should be smaller than the number of
independent measurements). Although $LU$ factorization would be one
alternative, perhaps a better choice is to use the Generalized
Singular Value Decomposition GSVD \cite{Hansen}, which allows the
regularized solution to be calculated efficiently for any value of $\alpha$.
The GSVD is now a standard tool for understanding the effect of the
choice of the regularization matrix $L$ in a linear ill-conditioned
problem, and has been applied to linearised EIT\cite{AndreaTMI}. The use of a
single linearized Tikhonov regularized solution is widespread in
medical industrial and geophysical EIT, the NOSER
 algorithm~\cite{NOSER}  being a well known example. Such algorithms
one step linear algorithms were the first three-dimensional algorithms
to be applied to experimental data from tanks~\cite{Goble}, and the
human thorax~\cite{Metherall,MetherallTh}. It must be emphasized that
a linearized solution will only be accurate when the true conductivity
is close to the initial estimate. 

\subsection{Backprojection}
It is an interesting historical observation that in the medical and
industrial applications of EIT numerous authors have calculated $J$
and then proceeded to use {\em ad hoc} regularized inversion 
methods to calculate an approximate solution. Often these are
variations on standard iterative methods which, if continued would for
a well posed problem converge to the Moore-Penrose generalised
solution. It is a standard method in inverse problems to use an
iterative method but stop short of convergence (Morozov's stopping
criteria tells us to stop when the output error first falls  below the
measurement noise). Many linear iterative schemes can be represented
as a filter on the singular values~\cite[Ch 1]{Vogel}. However they have the weakness that
the {\em a priori} information included is not as explicit as in
Tikhonov regularization. One extreme example of the use of an {\em ad
hoc} method is the method described by Kotre~\cite{Kotre} in which the
normalized transpose of the Jacobian is applied to the voltage
difference data. In the  Radon transform used in X-Ray CT
\cite{Natterer}, the formal adjoint of the Radon transform is called
the {\em back projection} operator. It produces at a point in the
domain the sum of all the values measured along rays through that
point. Although not an inverse to the Radon transform itself, a smooth
image can be obtained by backprojecting smoothed data, or equivalently
by back-projecting and then smoothing the resulting image. 

The Tikhonov
regularization formula \eref{eq:Tik} can be interpreted in a loose way
as the back-projection operator $J^*$ followed be the application of the spatial
 filter 
$(J^*J+ +\alpha^2 L^*L)^{-1}$. Although this approach is quite
different from the filtered back projection along equipotential lines of Barber
 and Brown~\cite{BB,SV} it is sometimes confused with this in the
literature. Kotre's back projection was until recently widely used in the
 process
tomography community for both resistivity (ERT) and permittivity (ECT)
imaging~\cite{WCIPT1}. Often
supported by the fallacious arguments, in particular that it is fast
(it is no faster than the application of any precomputed regularized
inverse) and that it is  commonly used (only by those who know
no better). In an interesting development the application of a
normalised adjoint to the residual voltage error for the linearised
problem was suggested for ECT, and later recognised as yet another
reinvention of the well known Landweber iterative method~\cite{Yang}. Although
there is no good reason to use pure linear iteration schemes directly
on problems with such small a number of parameters
as they can be applied much faster using the SVD, an interesting
variation is to use such a slowly converging linear solution together
with projection on to a constraint set. A method which has been shown
to work well in ECT \cite{Byars}.

\subsection{Iterative solutions}
The use of linear approximation is only valid for small deviations
from the reference conductivity. In medical problems conductivity
contrasts can be large, but there is a good case for using the
linearized method to calculate a change in admittivity between two
states, measured either at different times or with different
frequencies. Although this has been called ``dynamic imaging'' in EIT
the term {\em difference imaging} is now preferred ({\em dynamic imaging}
is a better used to  describe statistical time series method such as \cite{Jari}). In industrial ECT  modest  variations of permittivity are
commonplace. In industrial problems and in phantom  
tanks it is possible to measure a reference data set using a
homogenious tank. This can be used to calibrate the forward model,
in particular the contact impedance can be estimated~\cite{RobertLassi}. In an
{\em in vivo} measurement there is no such possibility and it may be that
the mismatch between the measured data and the predictions from the
forward model is dominated by the errors in electrode position, boundary
shape and contact impedance rather than interior conductivity. Until
these problems are overcome it is unlikely, in the author's opinion,
to be worth using iterative non-linear methods {\em in vivo} using
individual surface electrodes.
 Note however that such methods are in
routine use in geophysical problems
\cite{Zhang,Loke2D,Loke3D}. Computational complexity of both forward
solution and inversion of the linearized system meant that, although
iterative nonlinear algorithms had been implemented for simulated data
on modest meshes earlier~\cite{Liu} it was only in the mid 1990s that
affordable computers had sufficient floating point speed and memory to
handle sufficiently dense three-dimensional meshes to fit tank data adequately~\cite{PaiviBarcelona,PaiviTh}

The essence of
non-linear solution methods is to repeat the process of
calculating the Jacobian and solving a regularised linear
approximation. However a common way to explain this is to start with
the problem of minimizing $f$, which for a well chosen $G$ will have a
critical point which is the  minimum. At this minimum $\nabla
f(\mathbf{s})=\mathbf{0}$ which is a system of $N$ equations in $N$
unknowns which can be solved by multi-variable Newton-Raphson
method. In practice for noisy data there may not be an exact solution. The  Gauss-Newton approximation to this, which neglects terms
involving second derivatives of $F$, is a familiar Tikhonov formula
updating the $n$ th approximation to the conductivity parameters $\mathbf{s}_n$
\begin{equation} \label{eq:GN}\mathbf{s}_{n+1} =\mathbf{s}_n +(J_n^*J_n +\alpha^2
L^*L)^{-1}(J_n^*(\Vmeas-F(\mathbf{s}_n))+\alpha^2L^*L(\srefer-\mathbf{s}_n) \end{equation}
where $J_n$ is the Jacobian evaluated at $\mathbf{s}_n$, and care has
to be taken with signs . 
Notice that in this formula the Tikhonov parameter is held constant
throughout the iterations, by contrast Levenberg-Marquardt\cite{Marquardt} method applied to
$\nabla f=0$ would add a diagonal matrix $\lambda D$ in addition to
the regularization term $\alpha^2L^*L$ but would reduce $\lambda$ to
zero as a solution was approached. For an interpretation of $\lambda$
as a Lagrangian multiplier for an optimization constrained by a {\em
trust region} see~\cite[Ch 3]{Vogel}. Another variation on this family
of methods is, given an update direction from the Tikhonov formula, to
do an approximate {\em line search} to minimize $f$ in that
direction. Both methods are described in \cite[Ch 3]{Vogel}.

The parameterization of the conductivity can be much more specific
than voxel values or coefficients of smooth basis functions. One
example is to assume that the conductivity is piecewise constant on
smooth domains and reconstruct the shapes parameterized  by Fourier
series~\cite{Ville1,Ville2} or by level sets~\cite{Dorn}. For this and other model
based approaches the same family of smooth optimization techniques can
be used as for simpler parameterizations, although the Jacobian
calculation may be more involved.

\section{Jacobian calculations}

In optimization based methods it is often necessary to calculate the
derivative of the voltage measurements with respect to a conductivity
parameter. The complete matrix of partial derivatives of voltages with
respect to conductivity parameters is the {\em Jacobian} matrix,
sometimes in the medical and industrial EIT literature called the {\em
sensitivity matrix}, or the rows are called {\em sensitivity maps}. We will describe here the basic method for calculating this
efficiently with a minimal number of forward solutions. Let it be said
first that there are methods where the derivative is calculated only
once, although the forward solution is calculated repeatedly as the
conductivity is updated. This is the difference between {\em
Newton-Kantorovich} method and Newton's method. There are also {\em
Quasi-Newton} methods in which the Jacobian is updated approximately
from the forward solutions that have been made. Indeed this has been
used in geophysics~\cite{Loke3D}. It also worth pointing out that were
the conductivity is parameterized in a nonlinear way for example
using shapes of an anatomical model, the Jacobian with respect to
those new parameters can be calculated using the chain rule.

\subsection{Dissipated  power}
The derivation of the Jacobian is best understood in terms of the
dissipated power.
We take for simplicity a real conductivity $\sigma$. 
The potential $u$ satisfies 
\begin{equation}\label{eq:divsiggradu}
\nabla\cdot\sigma\nabla u=0
\end{equation} 
Using  the weak form of~\eref{eq:divsiggradu} or using Green's identity, for any $w$
\begin{equation}
\int_\Omega \sigma \nabla u \cdot \nabla w \,dV= \int_{\partial \Omega} w   \sigma
\frac{\partial u}{\partial n}\,dS
\end{equation}
Here $dV$ and $dS$ are volume and surface measures, $n$ the outward
normal vector and $\Omega$ the domain (the body). We use the complete
electrode model~\cite{CEM} with contact impedance $z_l$ which says that on electrode $E_l$
\begin{equation}\label{eq:CEM}
u = V_l -  z_l
\sigma \frac{\partial u}{\partial n}
\end{equation}
for a constant electrode voltage $V_l$, the total current on electrode $E_l$ is
\begin{equation}
\int_{E_l} \sigma \frac{\partial u}{\partial n} dS = I_l
\end{equation}
and $ \partial u/ \partial n = 0$ away from electrodes. For the special case $w=u$ we have
the power conservation formula,
\begin{equation}
\int_\Omega \sigma |\nabla u|^2\,dV = \int_{\partial \Omega} u \,  \sigma
\frac{\partial u}{\partial n}\, dS = \sum\limits_l \int_{E_l} \left( V_l - z_l
\sigma \frac{\partial u}{\partial n}\right) \sigma \frac{\partial u}{\partial n}
\,dS
\end{equation} hence
\begin{equation}
\int_\Omega \sigma |\nabla u|^2\,dV + \sum\limits_l\int_{E_l} z_l\left( \sigma
\frac{\partial u}{\partial n} \right)^2 = \sum\limits_l V_l I_l
\end{equation}
This simply states that the power input is dissipated either in the domain $\Omega$
or by the contact impedance layer under the electrodes.

\subsection{Standard formula for Jacobian}\label{sec:jac}
We now take perturbations $\sigma \rightarrow \sigma+\delta\sigma$, $u \rightarrow
u+\delta u$ and $V_l\rightarrow V_l+\delta V_l$, with the current in each electrode
$I_l$ held constant. We calculate the first order perturbation, and
argue as in~\cite{Calderon} that the terms we have neglected are
higher order in the $L^\infty$ norm on $\delta\sigma$. The details of
the calculation are given for the complete electrode model case
in~\cite{Nick}, the result is
\begin{equation}\label{eq:powerpurt}
\sum\limits_l I_l \delta V_l = - \int_\Omega \delta \sigma |\nabla
u|^2 \, dV
\end{equation}

This gives only the total change in power, to get the change in voltage on a
particular electrode $E_m$ when a current pattern is driven in some or all of the
other electrodes we simply solve for the special `measurement current pattern'
$\widetilde{I}^m_l = \delta_{lm}$. To emphasize the dependance of the potential on a
vector of electrode currents $\mathbf{I}=(I_1,\dots,I_L)$ we write $u(\mathbf{I})$.
The {hypothetical} measurement potential is $u(\mathbf{I}^m)$, by contrast the
potential for the $d$-th drive pattern is $u(\mathbf{I}^d)$.  Applying the power
perturbation formula~(\ref{eq:powerpurt}) to $u(\mathbf{I}^d)+u(\mathbf{I}^m)$ and
$u(\mathbf{I}^d)-u(\mathbf{I}^m)$ and then subtracting gives the familiar formula
\begin{equation}
\delta V_{dm} = - \int_\Omega \delta \sigma \nabla u(\mathbf{I}^d)\cdot \nabla
u(\mathbf{I}^m) \, dV
\end{equation}
Standard arguments based on series expansions of
operators~\cite{Calderon,BreckonTh} can be used to show that this is
indeed the   Fr\'{e}chet derivative for $\delta\sigma\in
L^\infty(\Omega)$, considerable care is needed to show that the
voltage data is Fr\'{e}chet differentiable in other normed spaces, such as those
needed to show that the total variation regularization scheme
works~\cite{WadeSenior}. For a finite dimensional subspace of
$L^\infty(\Omega)$ a proof of differentiability is given by~\cite{Kaipio}.

In the special case of the Sheffield adjacent pair drive, adjacent pair measurement
protocol, commonly used in two dimensional EIT we have potentials $u_i$ for the
$i$-th drive pair and voltage measurement $V_{ij}$ for a fixed
current $I$a across the $j$-th measurement pair 
\begin{equation}
\delta V_{ij} = -\frac{1}{I^2} \int_\Omega \delta \sigma \nabla u_i\cdot \nabla u_j
\, dV
\end{equation}
To calculate the Jacobian matrix one must choose a discretizarion of the
conductivity. The simplest case is to take the
conductivity to be piecewise constant on polyhedral domains such as voxels or
tetrahedral elements. Taking $\delta \sigma$ to be the characteristic function of
the $k$-th voxel we have for a fixed current pattern
\begin{equation}\label{eq:jac}
\frac{\partial V_{dm}}{\partial \sigma_k} =  - \int_\mathrm{voxel \, k}  \nabla
u(\mathbf{I}^d)\cdot \nabla u(\mathbf{I}^m) \, dV
\end{equation}
With the double indices $dm$ renumbered as a single index, these
functions form the elements of the Jacobian matrix $J$. For the case of a complex admittance one must repeat this calculation taking care to
use the real component of dissipated power $V_l\bar{I_l}$. Some EIT and capacitance
tomography systems use  a constant voltage source and in this case the change in
power of an increase in admittivity will have the opposite sign to the constant
current case.

Some iterative nonlinear reconstruction algorithms, such as nonlinear
Landweber, or non-linear conjugate gradient~\cite{Vogel,Tarantola} require the
evaluation of transpose (or adjoint in the complex case) of the Jacobian multiplied by a
vector $J^*z$. For problems where the Jacobian is very large it may be
undesirable to store the Jacobian and then apply its transpose to
$z$. Instead the block of $z_i$ corresponding to the $i$th current drive
is written as a distributed source on the measurement electrodes. A
forward solution is performed with this as the boundary current
pattern so that when this measurement field is combined with the field
for the drive pattern as \eref{eq:jac}, and this is accumulated to
give $J^*z$. For details of this applied to diffuse optical tomography
see~\cite{Arridge}, and for a general theory  of adjoint sources
see~\cite{Vogel}. For an example of application to EIT see for
example~\cite[Ch 4]{NickTh} and to  electromagnetic imaging~\cite{Dorn2}.

For fast calculation of the Jacobian using~\eref{eq:jac} one can
precompute the integrals of products of finite element basis functions
over elements. If non constant basis functions are used on elements,
or higher order elements used one could calculate the product of
gradients of FE basis functions at quadrature points in each
element. As this depends only on the geometry of the mesh and not on the
conductivity this can be precomputed unless one is using an adaptive
meshing strategy. The same data is used in assembling the FE system
matrix efficiently when the conductivity has changed but not the
geometry. It is these factors particularly which make current
commercial FEM software unsuitable for use in an efficient EIT solver.
While there are some details of efficient methods for Jacobian
calculation in the literature (\cite{Kaipio}), more specific
implementation details can often be found in theses, such as~\cite{YorkeyTh,BreckonTh, MarkoTh, PaiviTh, NickTh}.

\section{Inverse crimes and  common pitfalls}

The ill-posed nature of inverse problems means that any reconstruction
algorithm will have limitations on what images it can accurately
reconstruct and that the images degrade with noise in the data. When
developing a reconstruction algorithm it is usual to test it initially
on simulated data. Moreover the reconstruction algorithms typically
incorporates a forward solver. A natural first test is to use the same
forward model to generate simulated data with no simulated noise and
to then find to one's delight that the simulated conductivity can be
recovered fairly well, the only difficulties being if it violates the
{\em a priori} assumptions built into the reconstruction and the limitations
of floating point arithmetic. Failure of this basic test is used as a
diagnostic procedure for the program. 

\subsection{Use a different mesh}
If one is fortunate enough to
have a good data collection system and phantom, and someone skilled
enough to make some accurate measurements with the system one could
then progress to attempting to reconstruct images from experimental
data. However more often the next stage is to test
further with simulated data and it at this stage that one must take
care not to cheat and commit a so called ``inverse crime''~\cite[p133]{ColtonKress}. The best
practice is to use an independent forward model, at the very least in
the case of a finite element forward model one would use a much finer
mesh, and one which was not a strict refinement of the mesh used in
the forward solver in the reconstruction. It is an obvious point
perhaps but the simulated conductivity must be represented on this
mesh and unless we are explicitly using {\em a priori} information about
for example the location of the boundary of an anomaly the mesh used
for simulate data should not be `known' to the reconstruction program.

\subsection{Simulating noise}
It is necessary to study the effect of measurement error on the
reconstruction. The first reflex of the mathematically trained is to
use a pseudo random number generator to add independent Gaussian noise
to each measurement with an identical variance. There are some tightly
controlled laboratory situations where experimentalists can carefully
calibrate their apparatus to remove systematic error leaving only
random error from discretization of measurements and random effects
such as thermal noise. By averaging large numbers of measurements the
Central Limit Theorem means that independent identical Gaussian noise
is a good approximation to the true statistics of the data
error. However in electrical imaging we are rarely in this fortunate
situation. Even using phantom tanks there are many sources of error
which are hard to calibrate away. The statistical characterisation of
instrumentation error in EIT is a topic which requires considerable
further study. {\em In vivo} studies suffer further sources of error
including variable contact impedance, motion artifact and variable
surface geometry all of which produce correlated errors in the data.

Even simple simulation of discretization error requires some
understanding of the measurement system one has in mind. While the
data is discretized into a binary representation of the voltage with
a fixed precision by an analogue to digital converter (ADC), the input
to the ADC has already been scaled by an amplifier. In a
Sheffield-type adjacent pair drive system~\cite{SheffAPT} this scale factor is
determined by the position of the measurement pair relative to the
drive pair to make best use of the range of the ADC. Multiple drive
systems employ different strategies~\cite{ACT3,OXBACT3}. One
should at least add identically distributed noise to suitably scaled
measurements. Whatever scheme is chosen for simulating noise it should
be carefully described, often a phrase such as ``$5\%$ random noise
was added'' is used without saying if this is $5\%$ of the largest
voltage (which could completely destroy smaller measurements) or that
percentage of each measurement.

\subsection{Pseudo random numbers}
 Two small caveats about the use of pseudo random number
generators, such as the {\tt randn} function in Matlab. First that
unless the seed for the pseudo-random number generator is changed 
 the same sequence of random numbers is generated each time Matlab is
started. It possible therefore that many of us are using the same
sequence of ``random'' numbers! Secondly that for each error level one
should generate a sequence or ensemble of pseudo-random errors and perform the
reconstruction for each. One can then find the mean error in
reconstruction and the mean error in fitting the data, as well as their
variance. As the EIT reconstruction problem is nonlinear adding
Gaussian error to the data will not produce even multivariate Gaussian
reconstruction error, so strictly one would have to consider the
full distribution of the errors, but for small noise levels one can
expect a linear approximation to be valid. In practice the danger is
that the pseudo-random sequence will produce an out-lier, and  one
plots reconstruction error against data error one would notice
the discrepancy, which presumably explains why the culprits usually
get away with the ``inverse misdemeanour'' of not using an ensemble of
pseudo random errors.

\subsection{Thou shallst not tweak}
Finally there is some sharp practice which applies equally to simulated
and experimental results. The first of these we will call
``tweaking''. Reconstruction programs have a number of adjustable
parameters such as Tikhonov factors and  stopping criteria for
iteration, as well as levels of smoothing, basis constraints and small
variations on algorithms. While there are rational ways of choosing
reconstruction  parameters
based on the data (such as generalized cross validation and L-curve),
and on an estimate of the data error (Morotzov's stopping criterion),
a practical procedure often employed is to simulate, or measure
experimentally, a variety of conductivity distributions the find by
trial and error parameters in the reconstruction program which do 
reasonably well, knowing what the image was meant to be.  One could
compare this with the training set used for a neural net. We would
then not be surprised to find that conductivity distributions close to
the training set can be reconstructed satisfactorily. A more
interesting test is to see how the algorithm performs when we deviate
from the training set. There are plenty of examples of conference
papers where reconstruction algorithms are shown to perform very well on single
circular anomalies, while reconstructions of complex objects with
varying contrasts are absent. As we know the problem is illposed, it
is inevitable that there are some conductivity distributions which can
not be reconstructed well with a particular algorithm, in particular
ones which violate the {\em a priori} assumptions. It is therefore no
dishonour to present the failures as well as the successes.  To avoid
the temptation to tweak the algorithm to produce the best results
given that the correct result is known, the best procedure would be to
conduct blind trials for both simulated and experimental data.

The author is aware that what is suggested here  best practice is a
high standard to aim for, and that examples of this crimes and
misdemeanours can be found in his own work, but the intention is to
elevate the standard in published work in EIT reconstruction generally
and to highlight these pitfalls. While in conference presentations it
is acceptable to describe work in progress, and to confess to any
inverse crimes and misdemeanours, it is the author's opinion that they
should be avoided in journal publications.

\section{Further developments in reconstruction}

In this breif review there is no space to describe in any detail many
of the exciting current development in reconstruction
algorithms. Fortunately many of them are treated in other articles in
this special  issue. Before highlighting some of these developments it
is worth emphasising that for ill-posed problem {\em a priori}
information is essential for  stable reconstruction algorithms, and
it is better that this information is incorporated in the algorithm in
a systematic and transparent way. Another general priciple of inverse
problems is to think carefully what information is required by the end
user. Rather than attempting to produce an accurate image what is
often required in medical (and indeed most other) applications is an
estimate of a much smaller number of parameters which can be used for
diagnosis. For example we may know that a patient has two lungs as
well as other anatomical features but we might want to estimate their
water content to diagnose pulminary oedema. A sensible strategy would
be to devise an anatomical model of the thorax and fit a few parameters
of shape and conductivity rather than pixel conductivity values. The
disadvantage of this approach is that each application of EIT gives
rise to its own specialised reconstruction method, which must be
carefully designed for the purpose. In the author's opinion the future
development of EIT systems, including electrode arrays and data
acquisition systems as well as reconstruction software, should focus
increasingly on specific applications, although of course such systems
will share many common components.

\subsection{Beyond Tikhonov regularization}

We have also discussed the use of more general regularization
functionals including total variation. For smooth $G$ traditional
smooth optimization techniques can be used, whereas for total
variation the Primal Dual Interior Point Method is
promising~\cite{BorsicTh}. In general there is a trade-off between 
incorporating accurate {\em a priori} information and speed of
reconstruction. Where the regularization term is a partial
differential operator, the solution of the linearized problem is a
compact perturbation of a partial differential equation. This suggests
that multigrid methods may be used in the solution of the inverse
problem as well. For a single linearized step this has been done for
the EIT problem by McCormick and Wade~\cite{MccormickWade}, and for
the non-linear problem by Borcea~\cite{Borcea}. In the same vein
adaptive meshing can be used for the inverse problem~\cite{Marc}. In
both cases there is the interesting possibility to explore the
interaction between the meshes used for forward and inverse solution.

 At the extreme end of this spectrum we would like to describe
the prior probability distribution and for a known distribution of
measurement noise and calculate the entire posterior
distribution. Rather than giving one image, such as the MAP estimate, this
gives a complete description of the probability of any image. If the
probability is bimodal for example, one could present the two local
maximum probability images. If one needed a diagnosis, say of a
tumour, the posterior probability distribution could be used to
calculate the probability that a tumour like feature was there. The
computational complexity of calculating the posterior distribution for
all but the simplest distributions is enormous, however the posterior
distribution can be explored using the Markov Chain Monte Carlo Method
(MCMC)~\cite{Kaipio}. This was applied to simulated EIT
data~\cite{FoxNicholls},and more recently to tank data, including in
this special issue~\cite{MCMC}. For this to
be a viable technique for the 3D problem highly efficient forward
solution will be required, and an efficient and fast 3D MCMC algorithm
for EIT presents a serious but very worthwhile challenge.

\subsection{Direct nonlinear methods}
Iterative methods which use optimization methods to solve a
regularized problem are necessarily time consuming. The forward
problem must be solved repeatedly and the calculation of an updated
conductivity is also expensive. The first direct method to be proposed
was the Layer Stripping algorithm~\cite{layerstrip} however this is
yet to be shown to work well on noisy data. An exciting recent
development  is the implementation of a Scattering Transform algorithm
proposed by Nachman.  Siltanen~\cite{Siltanen} showed that
this can be implemented stably and applied to tank  data. The main
limitation of this technique is that is inherently two dimensional and
no one has found a way to extend it to three dimensions, also in
contrast to the more explicit forms of regularization it is not clear
what {\em a priori} information is incorporated in this method as the
smoothing is applied by filtering the data. A strength of the method
is its ability to accurately predict absolute conductivity levels. In
some cases where long electrodes can be used and the conductivity
varies slowly in the direction in which the electrodes are oriented a
two dimensional reconstruction may be a useful approximation. This is
perhaps more so in industrial problems such as monitoring flow in
pipes with ECT or ERT. In some situations a direct solution for a two
dimensional approximation could be used as a starting point for an
iterative three dimensional algorithm.

Two further direct methods show considerable promise for specific
applications. The monotonicity method of Tamburrino {\em et al}~\cite{Tamburrino} relies on the monotonicity of the map $\rho
\mapsto R_\rho$ where $\rho$ is the real resistivity and $R_\rho$ the
transfer impedance matrix. This method, which is extremely fast, relies
on the resistivity of the body to be known to be one of two values. It
works equally well in two and three dimensions and is robust in the
presence of noise. The time complexity scales linearly with the
number of voxels (which can be any shape) and scales cubically in the
number of electrodes. It works for purely real or imaginary
admittivity, (ERT or ECT), and for Magnetic Induction Tomography for
real conductivity. It is not known if it can be applied to the complex
case and it requires the voltage measurements on current carrying electrodes.

Linear sampling methods \cite{Bruhl,Schappel} have a similar time complexity and
advantages as the monotonicity method. While still applied to piecewise
constant conductivities, linear sampling methods can handle any number
of discrete conductivity values provided the anomalies are separated from
each other by the background. The method does not give an indication
of the conductivity level but rather locates the jump discontinuities
in conductivity.  Both monotonicity and linear sampling methods are likely to find
application in situations where a small anomaly is to be detected and located, for
example breast tumours. 

Finally a challenge remains to recover anisotropic conductivity which
arises in applications from fibrous or stratified media (such as muscle), flow of
non-spherical particles (such as red blood  cells), or from compression (for
example in soil). The inverse conductivity problem at low frequency is
known to suffer from insufficiency of data, but with sufficient prior
knowledge (for example \cite{Lionheart}) the uniqueness of solution can be
restored. One has to take care that the imposition of a finite element
mesh does not predetermine which of the family of consistent solutions
is found~\cite{Juan}.

\section{Conclusions}

In conclusion, until medical EIT data is reconstructed using the best
available methods the results will be inconclusive. Many an
experimental study has be spoilt when carefully collected data has
been attacked by a crude two-dimensional linear reconstruction
algorithm and the resulting `blurry blobs' taken as evidence that EIT
is not suitable for the desired task. Careful consideration of {\em a
priori} information, measurement error and the model parameters
required is needed, together with close collaboration between
mathematicians and experimentalists. Periodically a mood arises at an
EIT meeting that the technique will never find real application in
medicine, however it is my contention that a particular application
should not be dismissed as impossible until both hardware and software
specialists working together have given it their `best shot'.

\section{Acknowledgements}
The author would like to thank the referees for their helpful
comments, and the colleagues, especially Nick Polydorides, who
commented on earlier drafts.

\end{document}